\documentclass[mathleft,fleqn]{an}
\usepackage{amssymb}
\usepackage{graphicx}
\usepackage[varg]{txfonts}
\usepackage{natbib}
\bibpunct{(}{)}{;}{a}{}{,}
\begin{document}

% The following seven commands are intended for editorial usage and
% should be ignored by the author(s).
\Pagespan{1}{}% Document's page range. 
% If second parameter is left empty, the last page is computed
% automatically.
\Yearpublication{2016}%
\Yearsubmission{2016}%
\Month{0}%   
\Volume{999}%  
\Issue{0}% 
\DOI{asna.201600000}% 

%%%%%%%%%%%%%%%%%%%%%%%%%%%%%%%%%%%%%%%%%%%%%%%%

\title{Study of the variability of Nova V5668 Sgr, based on high resolution 
spectroscopic monitoring}
\titlerunning{Spectroscopic monitoring of Nova V5668 Sgr}

\author{D. Jack\inst{1}\fnmsep\thanks{Corresponding author: {dennis@astro.ugto.mx}}
        \and  J. de J. Robles P\'erez\inst{1}  \and
I. De Gennaro Aquino \inst{2} \and K.-P. Schr\"oder\inst{1}
\and  U. Wolter\inst{2} \and P. Eenens\inst{1} \and 
J. H. M. M. Schmitt\inst{2} \and M. Mittag\inst{2} \and A. Hempelmann\inst{2}
\and J. N. Gonz\'alez-P\'erez\inst{2} \and G. Rauw\inst{3} \and P. H. Hauschildt\inst{2}
}
\institute{
Departamento de Astronom\'\i{}a, Universidad de Guanajuato, A.P.~144, 
      36000 Guanajuato, GTO, Mexico
\and Hamburger Sternwarte, Universit\"at Hamburg, Gojenbergsweg 112, 
      21029 Hamburg, Germany
\and Groupe d'Astrophysique des Hautes Energies, Institut 
d'Astrophysique et de 
G{\'e}ophysique, Universit{\'e} de Li{\`e}ge, All{\'e}e du 6 Ao{\^ut},
B{\^a}t B5c, 4000 Li{\`e}ge, Belgium}

\authorrunning{D. Jack et al.}

\received{XXXX}
\accepted{XXXX}
\publonline{XXXX}
      
\label{firstpage}


% Abstract of the paper
\abstract{
We present results of our dense spectroscopic monitoring of Nova V5668 Sgr.
Starting on March 19 in 2015, only a few days after discovery,
we have obtained a series of spectra with the TIGRE telescope and its HEROS \'echelle
spectrograph which offers a resolution of $R\approx$ 20,000 and
covers the optical wavelength range from 3800 to 8800~\AA.
We performed a line identification of the discernible features for four spectra which are
representative for the respective phases in the light curve evolution of that nova. 
By simultaneously analysing the variations in the visual light curve
and the corresponding spectra of Nova V5668~Sgr,
we found that during the declining phases of the nova 
the absorption features in all hydrogen and many other lines had shifted to higher
expansion velocities of $\approx-2000$~km~s$^{-1}$.
Conversely, during the rise towards the following maximum, these
observed absorption features had returned to lower expansion velocities.
We found that the absorption features of some \ion{Fe}{ii} lines displayed the same behaviour,
but in addition disappeared for a few days during some declining phases.
Features of several \ion{N}{i} lines also disappeared while new \ion{N}{ii} lines appeared in emission for a few days
during some of the declining phases of the light curve of Nova V5668 Sgr. The
shape of the emission features is changing during the evolution and shows
a clear double peak structure after the deep minimum.
}

\maketitle

\section{Introduction}

A white dwarf (WD) in a close binary system can accrete hydrogen from a
Roche-lobe filling companion star. A layer of hydrogen accumulates in an accretion disk and, eventually,
on the degenerate surface of the WD.
At some point, the hydrogen ignites and a thermonuclear runaway starts and ejects
material from the surface of the WD. These explosive events are observed as classical novae
(see \citet{payne64,gallagher78,book08} or \citet{bode10} for reviews).
It is believed that the hydrogen keeps burning on the surface of the white dwarf,
as the detection of X-ray and its light curve indicate.
Nova envelopes are usually found to be close to the Eddington limit and
it is likely that this results in a continuous ejection of gas
comparable to a very strong stellar wind with the respective expansion
velocity profile.

To understand in more detail the physics of this explosive ejection of
gas in a nova event, dense spectroscopic monitoring of novae is an important tool,
since whatever is causing rapid changes in the light curve of a nova should also
have direct effects on its spectrum on the same very short time scales.
Performing the spectroscopic monitoring with high resolution is also desirable in order to observe
details in the respective nova absorption and emission features of the characteristic spectral lines.
\citet{novadel} studied the bright Nova V339 Del, which appeared in August 2013,
based on observations using spectroscopic monitoring with high resolution
obtained with the robotic telescope TIGRE
(Telescopio Internacional en Guanajuato, Rob\'otico y Espectrosc\'opico)
and presented a detailed atlas of lines that show features in the optical spectra of that nova.

On March 15 in 2015 a new and very bright classical nova appeared in the constellation
of Sagittarius \citep{ATel7230}. The Nova V5668 Sgr has been studied by many observers and
also in different wavelength regions.
Near infrared observations clearly show the formation of dust in this nova when
it drops to a deep minimum in the light curve after about 90 days after
discovery \citep{banerjee16, ATel7748}. 
Carbon monoxide has also been detected in its spectra \citep{ATel7303}.
During the later phases, Nova V5668 Sgr could also be observed in X-rays \citep{ATel7953,ATel8054,ATel8133}
as well as in gamma rays \citep{ATel7283,ATel7315}.
The \textit{Hubble Space Telescope} has observed the nova with observations performed in the ultraviolet
wavelength region using the STIS instrument \citep{ATel8275}.
Furthermore, polarimetry data have been taken of that nova \citep{ATel7986}.
The observations resumed once Nova V5668 Sgr was again observable
after its conjunction with the Sun showing that dust emission is still present but
has weakened over the months \citep{ATel8753}.

In this paper, we present the results of our analysis of the spectroscopic monitoring of Nova V5668 Sgr
with the TIGRE telescope.
We give the details of our observations and a short discussion of the light curve of the nova in Section \ref{sec:obs}.
The results that we obtained from our spectroscopic monitoring of the nova during the
different phases of the light curve evolution are shown in the following Section \ref{sec:mon}.
We will close this paper with Section \ref{sec:conclusion}, where we present a discussion and our conclusions.

\section{Observations of Nova V5668 Sgr}\label{sec:obs}

Starting from day March 19 UT in 2015 on, we observed a dense time series of spectra of 
Nova V5668~Sgr \citep{kazarovets15}
-- also known as Nova Sagittarii 2015 No.~2 or PNV J18365700-2855420 --
with the 1.2~m robotic telescope TIGRE situated
near Guanajuato, Central Mexico, and equipped with the \'echelle spectrograph HEROS
(Heidelberg Extended Range Optical Spectrograph).
The time series was interrupted on June 17,
when the nova was in a steep decline about $90$ days
after its discovery. Between March 19 and June 17 we were able to
obtain a total of 46 spectra during this period of 91 possible days of observation. 
Starting on July 26, 133 days after discovery, when the light curve of Nova V5668 Sgr
had recovered from the deep minimum, we resumed monitoring
until October 11 in 2015, 210 days after discovery,
when the nova had started to be too close to the Sun and was eventually heading for its conjunction.
During this phase we obtained 26 spectra, which gives a total of 72 spectra
of Nova V5668 Sgr.

With the \'echelle spectrograph HEROS, the spectra have a quite high
resolution of $R\approx 20,000$ and
cover the optical wavelength range from 3800 to 8800~\AA, with
just a small gap of about 130~\AA\ around 5800~\AA. The spectra are obtained in
two channels (red and blue) simultaneously.
For the observations we aimed for exposure times of one hour
to obtain a high signal-to-noise ratio of $>100$.
The spectra after the deep minimum have only a signal-to-noise ratio of $S/N\approx15$,
which is due to the lower brightness of the nova and the low continuum emission during that phase.
However, the emission lines in the spectra
show a signal-to-noise ratio that is significantly higher, therefore still allowing for studying these
features in detail.
Since TIGRE is a robotic telescope the observations as well as the data reduction
were performed automatically.
It needs to be stated that the observed spectra could not be calibrated for absolute
fluxes, but we were able to obtain all the spectra with relative fluxes.
For a more detailed technical description of the TIGRE instrumentation 
and its capabilities, see \citet{schmitt14}. 
The TIGRE telescope, although originally not designed for it, has turned out
to be very useful for spectroscopic monitoring of high energy astrophysical events
like novae \citep{novadel} and supernovae \citep{jack15b}.

Nova V5668~Sgr also has been observed with an even higher resolution of
$R\approx 60,000$ by \citet{tajitsu16}, but
they only obtained one spectrum on May 29 and, therefore, did not perform a dense spectroscopic monitoring,
which is the big advantage of our observation campaign.
Observations with the PEPSI spectrograph and a resolution of $R\approx 270,000$ have
also been reported \citep{wagner16}.

\subsection{Visual light curve}

\begin{figure}
\begin{center}
 \resizebox{\hsize}{!}{\includegraphics{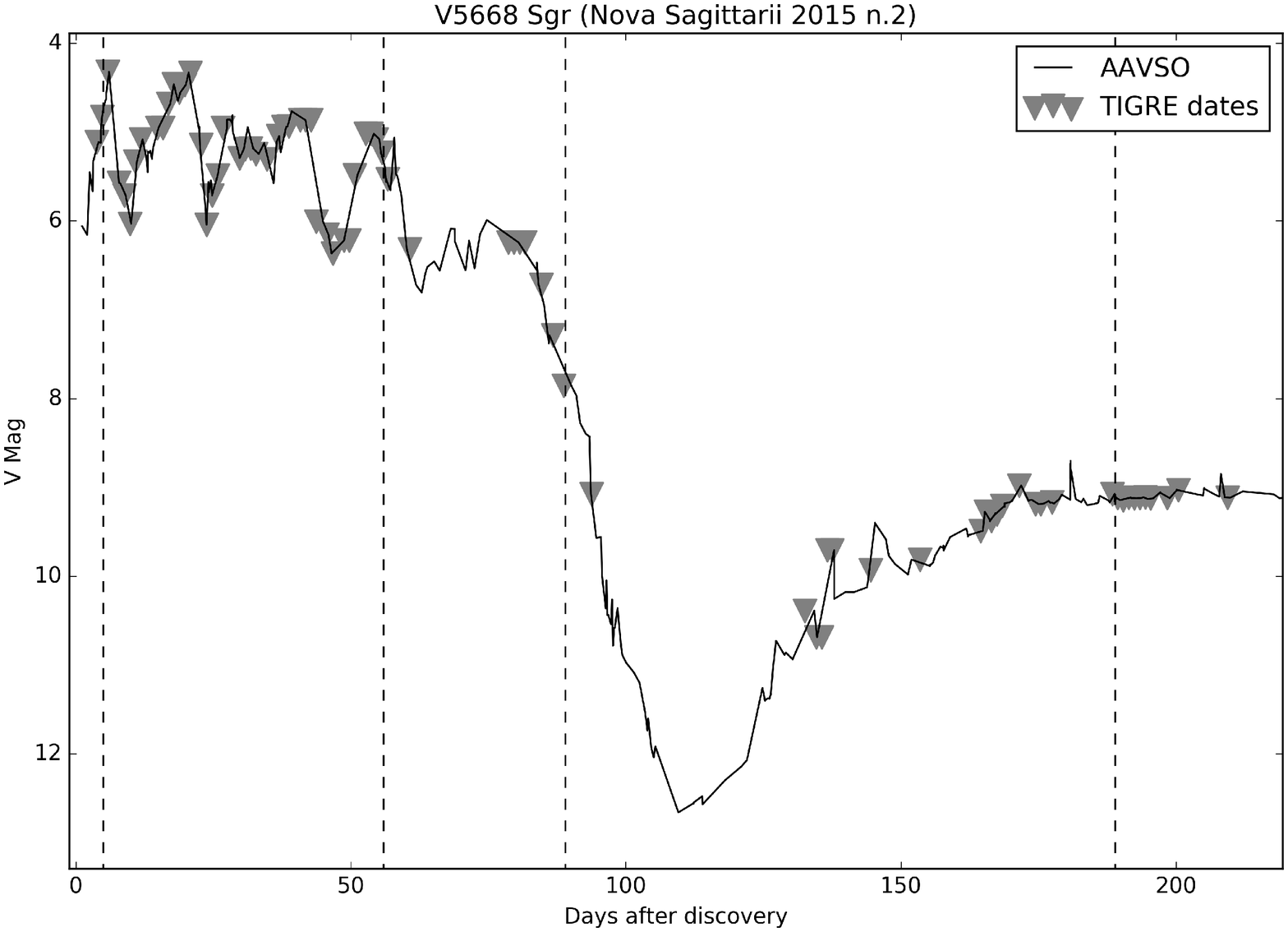}}
\caption{Visual light curve of Nova V5668~Sgr after its discovery on
March 15 in 2015. The data points were taken from AAVSO observations. The vertical lines
indicate the days where we performed a thorough line identification. The triangles
mark the days where we obtained TIGRE spectra.}
\label{fig:lc_tot}
\end{center}
\end{figure}

Fig. \ref{fig:lc_tot} demonstrates the visual light curve of Nova V5668~Sgr
starting from its day of discovery until the beginning of
November in 2015, when the nova started to be too close to the Sun for further
observations.
The data points were taken from the AAVSO Database\footnote{http://www.aavso.org},
and March 15 in 2015 was taken as day zero after the discovery.
The triangles mark the days where we obtained spectra with the TIGRE telescope.
As can be seen in the graph, during the first 90 days there
are significant variations in the light curve of up to 2~mag,
while the magnitude stays more or less between 5 and 7~mag.
This phase will be discussed in detail in the following Section \ref{sec:lc_var}.
After the variation phase, the light curve of Nova V5668 Sgr shows a very steep
decline down to a minimum with a magnitude of $\approx13.5$~mag.
After this deep minimum 
the visual light curve rose again up to a magnitude of $\approx9$~mag.
The light curve then stayed almost constant and subsequently declined very slowly.
The vertical lines in Fig. \ref{fig:lc_tot} mark the days after discovery where we performed
a detailed line identification in the observed spectra (see Section \ref{sec:ident}).

The observed visual magnitude during the first maximum of Nova V5668~Sgr
is about 4.2~mag. Using the accurate CCD measurements of the AAVSO Database,
the color in $B-V$ is found to be 0.26~mag during the first days.
During this phase the intrinsic $(B-V)_0$ due to an effective temperature of $\approx 9000$~K
is about $0.05$~mag. This gives an estimate for the interstellar
extinction of $E(B-V)=0.21$~mag, which corresponds to a moderate extinction
of $A_V=0.7$~mag. Assuming an absolute magnitude of $M_{\rm vis}\approx -7.5$~mag, like
the one observed for DQ Her, we obtain a distance modulus of $11.0$~mag,
which places the Nova V5668~Sgr at a distance of about 1.6~kpc.
\citet{banerjee16} determined the distance to Nova V5668~Sgr to have
a value of 1.54~kpc and state that this is in good agreement with the
MMRD estimate using the relation of \citet{dellavalle95}.

\subsection{Light curve variation phase}\label{sec:lc_var}

\begin{figure}
\begin{center}
 \resizebox{\hsize}{!}{\includegraphics{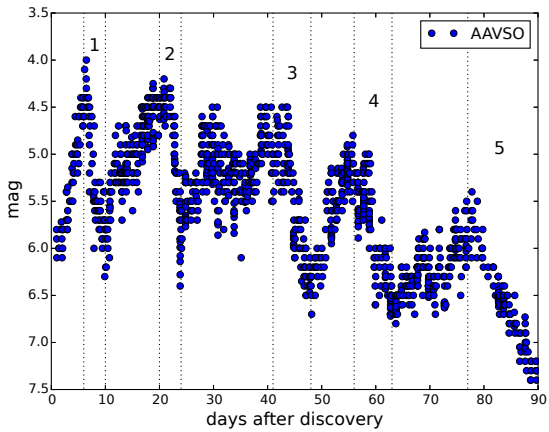}}
\caption{Visual light curve of Nova V5668~Sgr during the first 90 days after its discovery on
March 15 in 2015. The data points were taken from AAVSO observations. Five clear declining
phases have been identified.}
\label{fig:lc}
\end{center}
\end{figure}
In Fig. \ref{fig:lc}, the visual light curve of Nova V5668~Sgr is
shown for the first 90 days after discovery.
The data points were taken from AAVSO observations. 
As can be seen this part of the light curve shows five clear declining phases that are
indicated in the graph. These declines occur during just few days and
are also in general steeper than the following phases of increasing brightness.
We will study the corresponding changes that are observed in the optical spectra in the following chapter. 

In comparison with other novae, the light curve of Nova V5668~Sgr
displays the typical characteristics of the very bright 
Nova DQ Herculis from 1934 \citep{gaposchkin61}, with many variations during 
a long period and a final decline around 90 days after discovery.
After a deep minimum the light curve recovers to show then a very slow declining phase. 
See also \citet{strope10} for other novae of that type and with similar light curves.

\section{Spectroscopic monitoring of Nova V5668~Sgr}\label{sec:mon}

Our spectroscopic monitoring of Nova V5668~Sgr covers all different phases
of the light curve evolution as presented above.
In this section, we first present four of our high resolution spectra 
taken at very different stages of the outburst,
along with a detailed line identification. We then offer a detailed analysis 
of the spectra of Nova V5668~Sgr and their evolution until after the deep 
minimum of the light curve, when all lines are only seen in emission.

\subsection{High resolution spectroscopic monitoring}

\begin{figure*}
\begin{center}
\includegraphics[width=\textwidth]{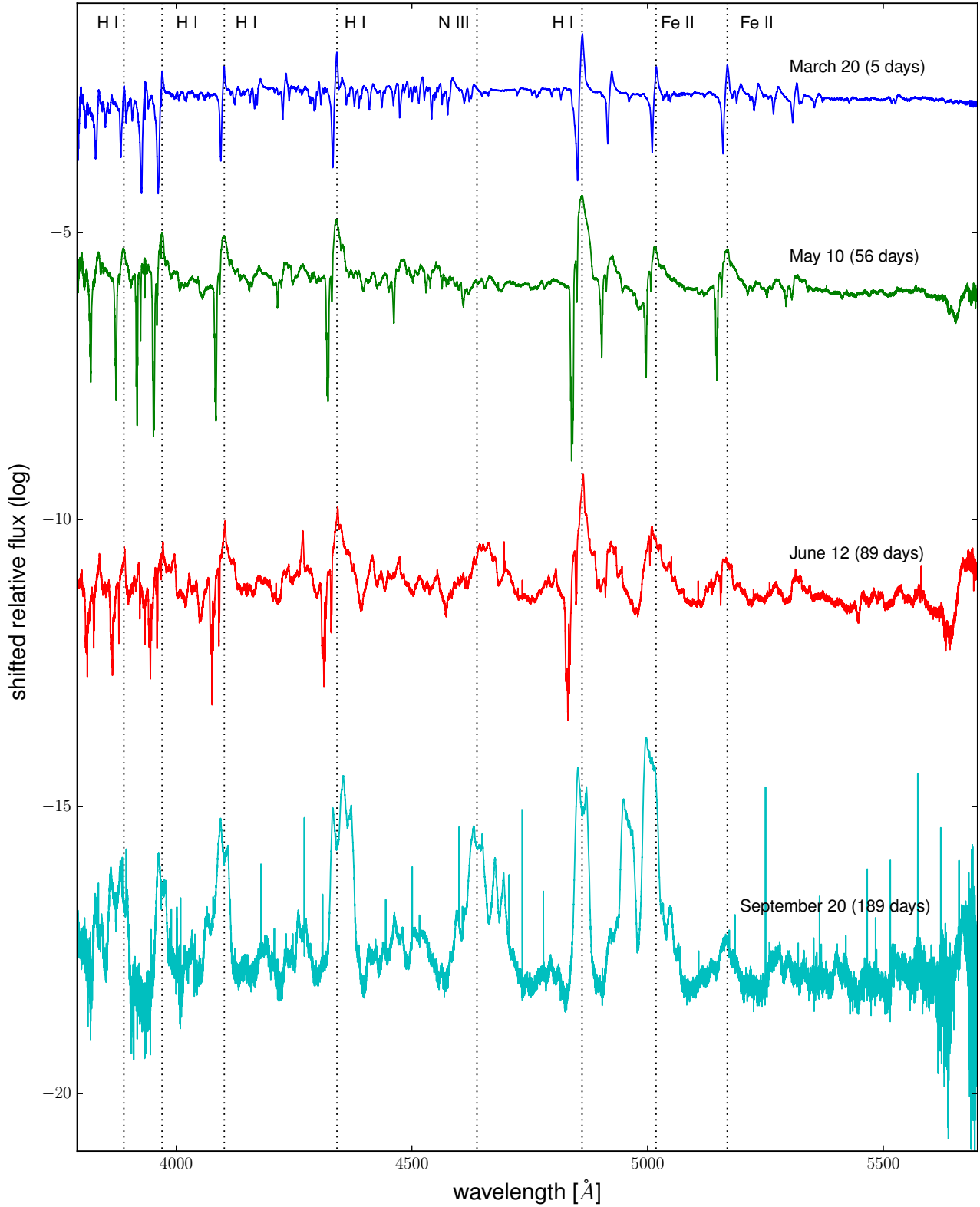}
\caption{Four spectra of Nova V5668 Sgr observed in the blue channel
of the HEROS spectrograph. Dates are of 2015 and the corresponding days
after discovery are given in brackets.}
\label{fig:all_specs_B}
\end{center}
\end{figure*}

\begin{figure*}
\begin{center}
\includegraphics[width=\textwidth]{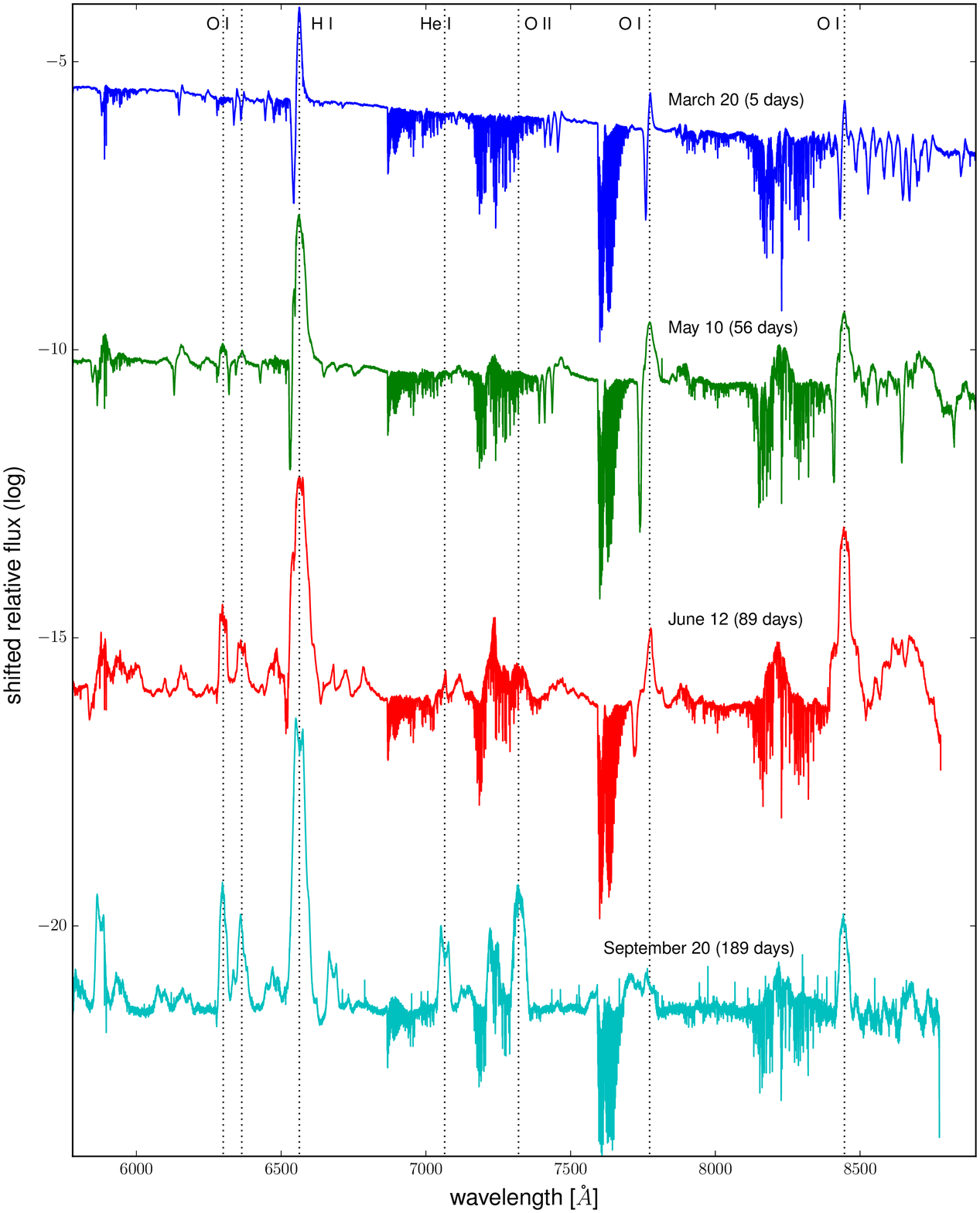}
\caption{Four spectra of Nova V5668 Sgr observed in the red channel
of the HEROS spectrograph. Dates are of 2015 and the corresponding days
after discovery are given in brackets. Broad bands of telluric lines are present in
this wavelength range.}
\label{fig:all_specs_R}
\end{center}
\end{figure*}

Fig. \ref{fig:all_specs_B} shows four spectra of the 
nova observed in the blue channel of the HEROS spectrograph between 3800 and 5800 \AA.
The spectra were taken at the dates indicated in the graph.
To represent the different phases of the nova, we chose a spectrum during
the first maximum as well as two spectra during the variation phase: one during a maximum, the other
when the light curve was declining towards the big minimum. The fourth spectrum spectrum
was taken after the minimum during the emission line or dust phase.
In Fig. \ref{fig:lc_tot}, the corresponding days
after discovery are marked by vertical lines.

The first spectrum presented was taken on March 20 in 2015, which corresponds to 5 days after discovery,
when the nova was around its maximum brightness.
The spectrum shows the characteristic P-Cygni profile features which one expects during this phase.
We marked some of the strongest and most important features in the spectrum with vertical
lines at their respective rest-wavelengths. All features show the emission peak 
at around the rest-wavelength and a blue shifted absorption feature corresponding to a
negative expansion velocity caused by the ejected gas of the nova event that is moving
towards the observer.
Features of all the Hydrogen Balmer lines that can be observed in this wavelength range are clearly visible
(H~$\beta$ at 4861~\AA, H~$\gamma$ at 4341~\AA, H~$\delta$ at 4102~\AA, etc.).
Two clear \ion{Fe}{ii} features are also marked in Fig. \ref{fig:all_specs_B}, but there are many more iron
lines present in that  spectrum (see Section \ref{sec:ident} for a detailed list of identified lines).
The second spectrum presented was taken on May 10 during the light curve variation phase around the
fourth maximum 56 days after discovery. 
All Balmer lines are still present, but the line profiles have already changed.
The emission features are now broader, and the absorption features are still present.
Looking closely one can see that there are now two distinct absorption features in most of the lines.
On June 12 we took the third spectrum presented here. It was taken 89 days after discovery
when the light curve was in the steep decline towards the deep minimum.
The Balmer lines show mainly emission features but still have some
absorption features, while the marked features of the \ion{Fe}{ii} lines are only in emission. 
In the last presented spectrum, which was observed after the deep minimum 189 days after discovery
on September 20, no absorption features were found, but some new emission
lines appear, such as one at 4638.0~\AA, presumably due to \ion{N}{iii},  
as can be seen in Fig.~\ref{fig:all_specs_B}.

In Fig.~\ref{fig:all_specs_R}, we present four spectra of Nova V5668 Sgr observed by
the TIGRE telescope and in the red channel of
the HEROS spectrograph in the wavelength range from 5800 to 8800~\AA.
The spectra were taken at the same days as the spectra presented in Fig. \ref{fig:all_specs_B}
and as indicated by the vertical lines in Fig. \ref{fig:lc_tot}.
Comparing the spectra of the two channels, it can be seen that the spectra of the
red channel show less features than the ones that were taken in
the blue channel. The observations in the red channel are affected by telluric lines, which we did not remove in our
spectra.
In the spectra, the telluric lines are visible in Fig.~\ref{fig:all_specs_R}
around the wavelength ranges from 6800 to 7300~\AA, 7600 to 7700~\AA\ and 8000 to 8400~\AA.
The first spectrum in Fig.~\ref{fig:all_specs_R} from March 20 shows 
features of lines with clear P-Cygni profiles as was already observed in the blue channel.
Strong lines are H~$\alpha$ at a wavelength of 6562.7~\AA\ and an \ion{O}{i} line at 7773.0~\AA.
The lines in the second spectrum from May 10 still show P-Cygni profiles but the emission features are now broader
than before. In the H~$\alpha$ line has appeared an additional small absorption feature that
can also be seen in the two strong lines of \ion{O}{i} marked in the spectrum. In the third
spectrum from June 12, the line profiles have started to change to present now mainly emission features, but some lines
still show absorption features. Two new clear emission features of [\ion{O}{i}] at wavelengths of 6300.3 and 6363.7~\AA\
appear in the spectrum and are still visible in the fourth presented spectrum from September 20.
There also arises a new emission feature of the [\ion{O}{ii}] line at 7320.0~\AA,
which is clearly visible in this last spectrum.
In the spectrum from June 12 there are too many telluric lines in that wavelength region so that a
clear identification is difficult. A further new strong emission feature in this spectrum
is the \ion{He}{i} line at 7065.2~\AA, which is also present in the spectrum of Nova V5668 Sgr from September 20,
which was observed after the deep minimum 189 days after discovery.
As one can clearly see many of the emission features have a double peak shape.

\subsubsection{Line Identification}\label{sec:ident}

% Table of features
\begin{table}
\caption{List of lines in the blue channel that have been identified in four selected spectra of Nova V5668~Sgr
during characteristic light curve phases.}\label{tab:lines}
\centering
\begin{tabular}{cccccc}
\hline
Line [\AA] & Element & Mar 20 & May 10 & Jun 12 & Sep 20\\
\hline 
3835.4 & \ion{H}{i} & \checkmark  & \checkmark & \checkmark & \checkmark \\
3856.0 & \ion{Si}{i}  & \checkmark & & & \\
3862.6 & \ion{Si}{i} & \checkmark & & & \\
3889.1 & \ion{H}{i} &  \checkmark & \checkmark  & \checkmark & \\
3900.5 & \ion{Ti}{i} & \checkmark  & &  \\
3906.0 & \ion{Fe}{ii} &  \checkmark &  &  &  \\
3913.5 & \ion{Ti}{ii}  &  \checkmark & &  &  \\
3933.7 & \ion{Ca}{ii}  &  \checkmark & \checkmark & \checkmark &  \\
3968.5 & \ion{Ca}{ii}  &  \checkmark & \checkmark &  & \checkmark \\
3970.1 & \ion{H}{i}  & \checkmark & \checkmark & \checkmark & \\
3995.0 & \ion{N}{ii}  &  &  & \checkmark & \\
4077.7 & \ion{Sr}{ii} &  &  & \checkmark &  \\
4101.7 & \ion{H}{i} &  \checkmark & \checkmark &  \checkmark & \checkmark\\
4130.9 & \ion{Si}{ii}  &  \checkmark & \checkmark &  & \\
4173.5 & \ion{Fe}{ii}  &  \checkmark &  &  & \\
4178.9 & \ion{Fe}{ii}  &  \checkmark &  &  & \\
4233.2 & \ion{Fe}{ii}  &  \checkmark & \checkmark & \checkmark & \checkmark \\
4267.2 & \ion{C}{ii}  &   &  & \checkmark &  \\
4303.2 & \ion{Fe}{ii}  & \checkmark & \checkmark &  &  \\
4340.5 & \ion{H}{i}  & \checkmark & \checkmark & \checkmark & \checkmark\\
4351.8 & \ion{Fe}{ii}  & \checkmark & \checkmark & & \\
4385.4 & \ion{Fe}{ii}  & \checkmark &  & & \\
4416.3 & \ion{Fe}{ii}  & \checkmark &  & & \\
4416.8 & \ion{Fe}{ii}  &  &\checkmark  & \checkmark & \\
4471.5 & \ion{He}{i}  &  &  & \checkmark & \\
4481.2 & \ion{Mg}{ii}  & \checkmark & \checkmark & \checkmark & \\
4491.4 & \ion{Fe}{ii}  & \checkmark & \checkmark &  & \\
4514.9 & \ion{N}{iii}  &  &  & \checkmark & \\
4522.6 & \ion{Fe}{ii}  & \checkmark &  &  & \\
4549.5 & \ion{Fe}{ii}  & \checkmark &  &  & \\
4555.8 & \ion{Fe}{ii}  & \checkmark & \checkmark &  & \\
4572.0 & \ion{Ti}{ii}  & \checkmark & \checkmark &  & \\
4583.8 & \ion{Fe}{ii}  & \checkmark & \checkmark &  & \\
4629.3 & \ion{Fe}{ii}  & \checkmark & \checkmark &  & \\
4638.0 & \ion{N}{iii}  &   & & \checkmark & \checkmark \\
4685.8 & \ion{He}{ii}  & &  &  & \checkmark \\
4824.1 & \ion{Cr}{ii}  & \checkmark & & & \\
4861.3 & \ion{H}{i}  & \checkmark & \checkmark & \checkmark & \checkmark \\
4923.9 & \ion{Fe}{ii}  & \checkmark & \checkmark & \checkmark &  \\
4923.9 & \ion{Fe}{ii}  & \checkmark & \checkmark & \checkmark &  \\
5006.8 & \ion{O}{iii}  &  & & \checkmark & \checkmark \\
5018.4 & \ion{Fe}{ii}  & \checkmark  & \checkmark  & & \\
5045.1 & \ion{N}{ii}  & \checkmark  &   & & \\
5169.0 & \ion{Fe}{ii}  & \checkmark  & \checkmark  & & \\
5197.6 & \ion{Fe}{ii}  & \checkmark  &   & & \\
5234.6 & \ion{Fe}{ii}  & \checkmark  &  \checkmark & & \\
5276.0 & \ion{Fe}{ii}  & \checkmark  &  \checkmark & & \\
5316.6 & \ion{Fe}{ii}  & \checkmark  &  \checkmark & & \\
5362.8 & \ion{Fe}{ii}  & \checkmark  &   & & \\
5532.1 & \ion{Fe}{ii}  & \checkmark  &   & & \\
\hline
\end{tabular}
\end{table}

% second table
\begin{table}
\caption{List of lines in the red channel that have been identified in four selected spectra of Nova V5668~Sgr
during characteristic light curve phases.}\label{tab:lines2}
\centering
\begin{tabular}{cccccc}
\hline
Line [\AA] & Species & Mar 20 & May 10 & Jun 12 & Sep 20\\
\hline
5875.6 & \ion{He}{i}  &   &   &  \checkmark & \\
5889.9 & \ion{Na}{i}  &   \checkmark &  \checkmark &  \checkmark & \checkmark\\
5895.9 & \ion{Na}{i}  &   \checkmark &  \checkmark &  \checkmark & \checkmark\\
6247.6 & \ion{Fe}{ii}  &   \checkmark &   &   & \\
6300.3 & [\ion{O}{i}]  &    & \checkmark  & \checkmark  & \checkmark \\
6347.1 & \ion{Si}{ii}  &   \checkmark &   &   & \\
6363.7 & [\ion{O}{i}]  &   &  &   & \checkmark\\
6371.4 & \ion{Si}{ii}  & \checkmark  &  &   & \\
6456.4 & \ion{Fe}{ii}  & \checkmark  & \checkmark &   & \\
6562.7 & \ion{H}{i}  & \checkmark  & \checkmark &  \checkmark & \checkmark \\
7065.2 & \ion{He}{i}  &  & &  \checkmark & \checkmark \\
7115.0 & \ion{C}{i}  &  & \checkmark &   &  \\
7320.0 & [\ion{O}{ii}] &  &  &   & \checkmark \\
7442.3 & \ion{N}{i} &  \checkmark  &  &   & \\
7468.2 & \ion{N}{i} &    & \checkmark &   & \\
7773.0 & \ion{O}{i} & \checkmark  & \checkmark & \checkmark & \\
8446.3 & \ion{O}{i} & \checkmark  & \checkmark & \checkmark & \checkmark\\
8498.0 & \ion{Ca}{ii} & \checkmark  & \checkmark &  & \\
8542.1 & \ion{Ca}{ii} & \checkmark  & \checkmark &  & \\
8598.4 & \ion{H}{i} & \checkmark  &  &  & \\
8629.2 & \ion{H}{i} & \checkmark  &  &  & \\
8750.5 & \ion{H}{i} & \checkmark  &  &  & \\
8862.8 & \ion{H}{i} & \checkmark  & \checkmark & &\\

\hline
\end{tabular}
\end{table}

The four selected characteristic spectra described above were used to
identify all the lines that show spectral features.
For this thorough analysis we made use of the spectral atlas presented in \citet{novadel}
which was obtained from observations with the same telescope using spectra of Nova V339 Del.
In the Tables \ref{tab:lines} for the blue and \ref{tab:lines2} for the red channel, we list all the identified lines for which we found
features in the respective spectra. The tables are sorted by the rest-wavelength of the lines,
and it contains also the element and ionization stage to which the line belongs to.
Tick-marks indicate, in which of the four spectra the presence of a feature was clearly identified.

The first spectrum around the first maximum brightness contains the largest number of
lines with features in Nova V5668 Sgr.
We found many Balmer and also Paschen lines of Hydrogen as well
as \ion{Fe}{ii} lines and strong lines of \ion{O}{i} in the spectra of the nova.
Additionally, we could identify some lines of \ion{Ti}{ii}, \ion{Mg}{ii}, \ion{Si}{i}, \ion{Si}{ii},
\ion{N}{ii}, \ion{Na}{i} among others.
The hydrogen lines are present in all of the four spectra.
Other lines like some of \ion{Fe}{ii} could only be observed in the first spectra of Table~\ref{tab:lines}
and then disappeared in the later spectra.
There are also lines which were not observed in the first spectra,
but appeared in the later ones. These are some lines of \ion{He}{ii}, \ion{O}{iii}, \ion{N}{iii} among others.
We could also identify some forbidden lines of oxygen ([\ion{O}{i}] and [\ion{O}{ii}]), which
appear in the later spectra when the nova is in its optically thin nebular phase
after the deep minimum.
There might be also forbidden lines of nitrogen present in the spectra of Nova V5668 Sgr.
These lines are [\ion{N}{ii}] at 6548 \AA\ and [\ion{N}{ii}] at 6583\ AA.
However, the possible features of these lines coincide with the broad emission
feature of H$\alpha$ and it is difficult to clearly identify them. Fig. \ref{fig:Halpha}
might indicate the possible presence of these [\ion{N}{ii}] lines.

\subsection{The light curve variation phase}

During the light curve variation phase, we have taken optical spectra of Nova V5668~Sgr between 4 days and
94 days after discovery covering all five clear declining phases as indicated in Fig. \ref{fig:lc}.
This dense spectroscopic time series enables us to relate the spectral evolution directly
to the changes seen in the visual light curve and may reveal the physical circumstances
during this phase of small variations in the light curve.

A general observation during all of the declining phases of the nova is that when the light curve decreases,
the continuum flux decreases as well.
The same occurs respectively during the rising phases of the light curve namely the continuum flux increases.
Although we do not have flux calibrated spectra, we find that during declining phases
the emission line flux increases relative to the continuum. 
This rules out the possibility that the changes in the light curve are only due to changes in the emission line profiles
and is consistent with the constant colour in the AAVSO light curve of Nova V5668~Sgr.
In addition, we observed that some spectral features undergo significant changes as well. 
These allow the study of the physical properties of the envelope during 
variations in the visual light curve. It needs to be emphasised that this is possible only due to 
the dense time series of high resolution
spectra, which we obtained with the TIGRE telescope.

Regarding the spectra observed during the first maximum in the visual light curve,
before the maximum, the lines in the observed spectra show
clear P~Cygni profile features, which are 
caused by an optically thick expanding envelope, as it is the case for a nova.
After the first maximum the nova seems to change into its transition phase.
The spectra are now showing more complex features which are also beginning to change
into emission.
This is the typical behaviour of a classical nova, and the first maximum
of the light curve of Nova V5668~Sgr can be understood in this normal
context as well. However, one (and only one) forbidden line of \ion{O}{i} already appears in the spectra of 
the nova at that time (see Sect. \ref{sec:forbid}).

As is typical for DQ Her type novae, the visual light curve rises again after the first declining phase.
In our time series of spectra, we see that when the nova brightness declines steeply after its 
second maximum, the absorption features shift to notably higher
expansion velocities in several lines. Table \ref{tb:list} gives 
a list of lines which have features that show clearly this behaviour during the various 
declining phases towards the following minima.
The \ion{Na}{i} D doublet also shows the high expansion velocity absorption trough, 
but it needs to be stated that the two
individual lines are overlapping and that there is also a \ion{He}{i} line at 5875.6~\AA\ in the same
wavelength range that could have been blended into this features.
There are other lines, probably more \ion{Fe}{ii} lines, that showed this behaviour,
but it was not clearly seen in our spectra because of a low signal to noise or other reasons.
We listed only the lines where the behaviour could definitely be seen. 

% Table of features
\begin{table}
\caption{List of lines that show absorption features moving to higher
expansion velocities during the second to fifth declining phase
of the visual light curve of Nova V5668~Sgr.}
\centering
\begin{tabular}{cccccc}
\hline
Species & Line & 2. & 3. & 4. & 5.\\
\hline 
\ion{H}{i} & 3835.4 \AA  &   & \checkmark & \checkmark & \checkmark \\
\ion{H}{i} & 3889.1 \AA  &  \checkmark & \checkmark  & \checkmark & \checkmark \\
\ion{Ca}{ii} & 3933.7 \AA  & \checkmark &   &  & \\
\ion{Ca}{ii} & 3968.5 \AA  & \checkmark & \checkmark  & \checkmark & \checkmark \\
\ion{H}{i} & 4101.7 \AA  &  \checkmark & \checkmark & \checkmark & \checkmark \\
\ion{H}{i} & 4340.5 \AA  &  \checkmark & \checkmark & \checkmark & \checkmark \\
\ion{H}{i} & 4861.3 \AA  &  \checkmark & \checkmark & \checkmark & \checkmark \\
\ion{Fe}{ii} & 4923.9 \AA  &  \checkmark &  &  &  \\
\ion{Na}{i} & 5895.9 \AA  &  \checkmark & \checkmark & \checkmark & \checkmark \\
\ion{H}{i} & 6562.7 \AA  &  & \checkmark & \checkmark & \checkmark \\
\ion{He}{i} & 6678.2 \AA  &  \checkmark  & \checkmark & \checkmark & \checkmark \\
\ion{He}{i} & 7065.2 \AA  &  \checkmark & \checkmark & \checkmark & \checkmark \\
\ion{O}{i} & 7773.0 \AA  &  \checkmark & \checkmark &  \checkmark & \checkmark\\
\ion{H}{i} & 8862.8 \AA  &  \checkmark &  &  &\\
\hline
\end{tabular}
\label{tb:list}
\end{table}

% Figure
\begin{figure}
\begin{center}
 \resizebox{\hsize}{!}{\includegraphics{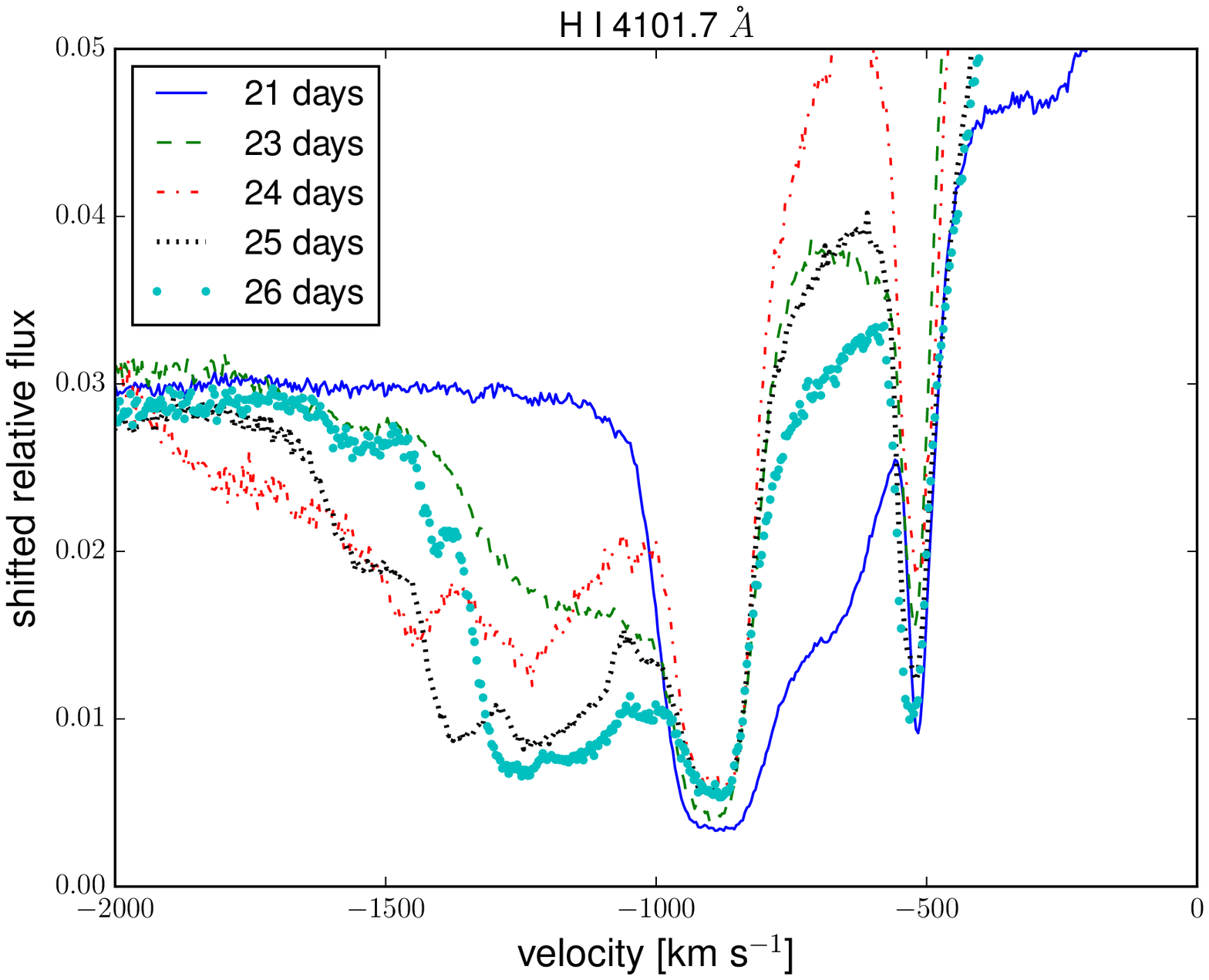}}
\caption{Evolution of the \ion{H}{i} line at a wavelength of 4101.7~\AA\ during
the second minimum phase observed in the visual light curve of Nova V5668~Sgr.
The spectra are shown at different days after discovery.}
\label{fig:HI}
\end{center}
\end{figure}

As an example of the different lines, which show the described behaviour, we present in Fig.~\ref{fig:HI}
the changes observed in the H~$\delta$ line at a wavelength of 4101.7~\AA. We concentrate on the negative
expansion velocity part of the line profile and ignore the emission part of the line which
has its maximum around the rest wavelength at 0~km~s$^{-1}$. 
21 days after discovery the light curve is still in its secondary maximum, and
the spectrum displays only absorption features below an expansion velocity of $\approx -1000$~km~s$^{-1}$.
These absorption features then shift to expansion velocities between
$-1500$ and $-1000$~km~s$^{-1}$
and during the minimum of the light curve 24 days after discovery,
these features reach their maximum expansion velocities of over
$-1500$~km~s$^{-1}$. During the following rise, they shift slowly towards
slower expansion velocities. 
The absorption feature at $\approx -500$~km~s$^{-1}$ can also be observed in many different lines.
However, it does not seem to change its position during the light curve variations.

\begin{figure}
\begin{center}
 \resizebox{\hsize}{!}{\includegraphics{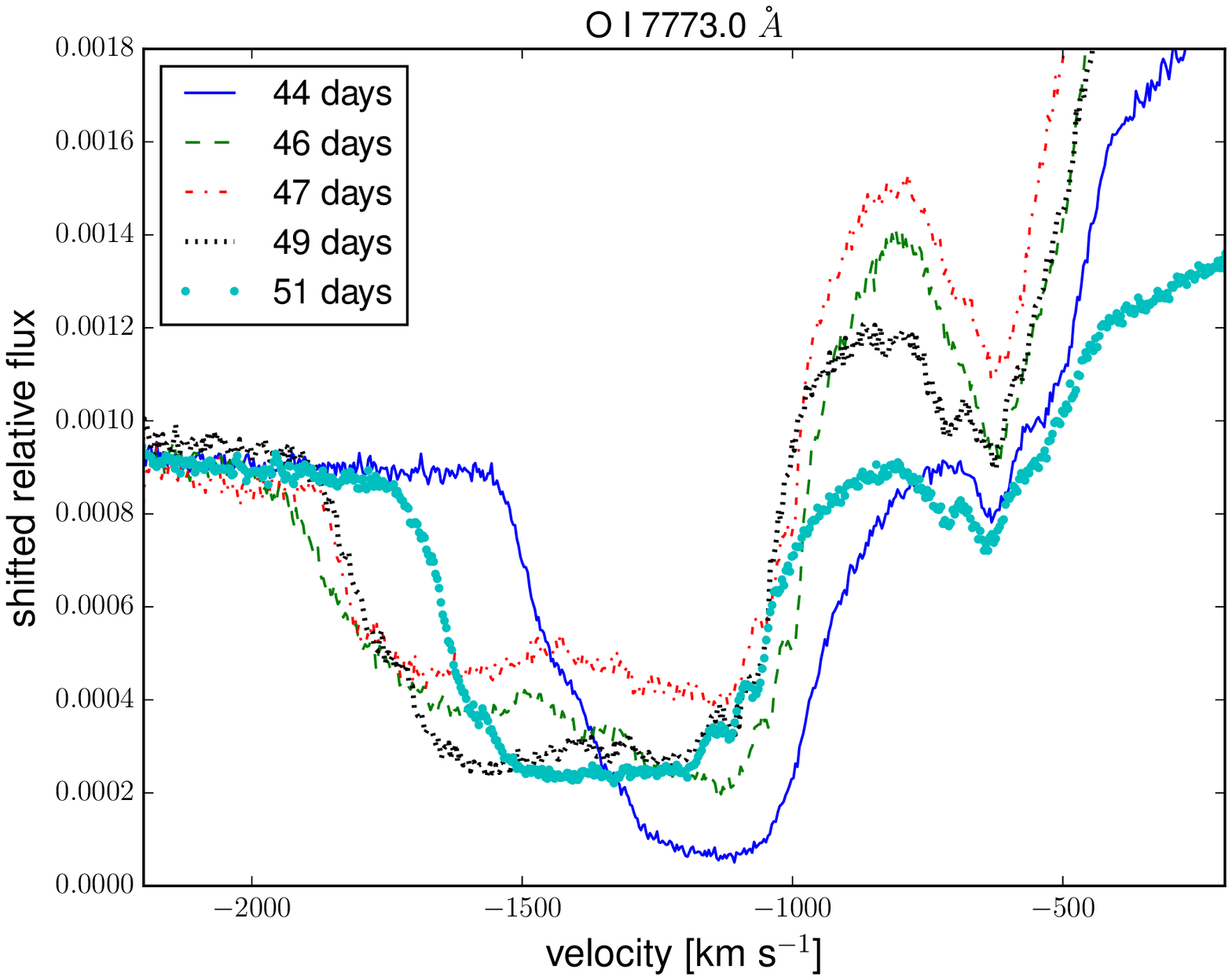}}
\caption{Evolution of the \ion{O}{i} line at a wavelength of 7773.0~\AA\ during
the third minimum phase observed in the visual light curve of Nova V5668~Sgr.
The spectra are shown at different days after discovery.}
\label{fig:OI}
\end{center}
\end{figure}

As a representative example for the third declining phase,
we show in Fig. \ref{fig:OI} the spectral evolution of the \ion{O}{i}
line at 7773.0~\AA. 
44 days after discovery, during the third maximum,
absorption features between $-1500$ and $-1000$~km~s$^{-1}$ were observed.
During the following minimum 46 days after discovery, absorption features
appear at expansion velocities between $-2000$ and $-1500$~km~s$^{-1}$ and
shift during the following rise to slower expansion velocities
as can be seen in the spectra of 49 and 51 days after discovery.

During the fourth declining phase we observed the nova
56 and 57 days after discovery, which is during the maximum phase.
Due to bad weather and scheduled maintenance of the
telescope we obtained only one spectrum during the declining phase 61 days after discovery.
However, we could again observe that high velocity absorption features appear
in some of the lines, this time at relatively high expansion
velocities of between $-2500$ and $-2000$~km~s$^{-1}$.
Unfortunately, we cannot say anything about the 
subsequent behaviour of these features since we did not cover the following 
rising phase of the visual light curve.
The observed expansion velocities of the absorption features during the fifth declining phase
move to between $-2100$ and $-1700$~km~s$^{-1}$. This fifth declining phase has
eventually a steep decline down to the deep minimum in the light curve
of Nova V5668 Sgr.

\begin{figure}
\begin{center}
 \resizebox{\hsize}{!}{\includegraphics{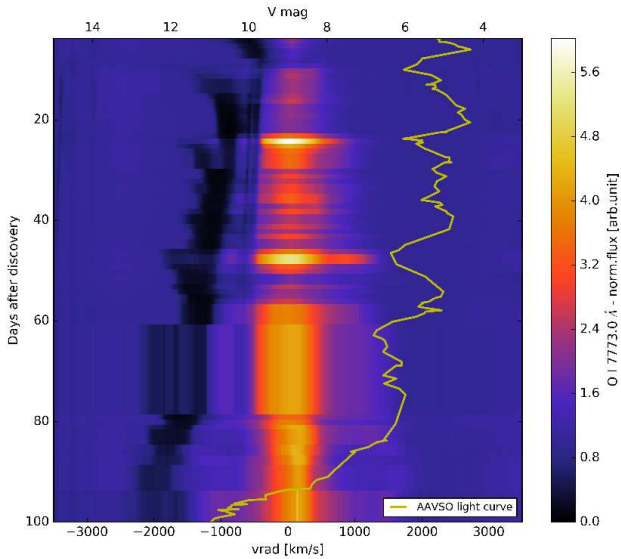}}
\caption{Evolution of the \ion{O}{i} line profile at a wavelength of 7773.0~\AA\ during
the light curve variation phase of the visual light curve (continuous line) of Nova V5668~Sgr
during the first 100 days after discovery.}
\label{fig:OI_2d}
\end{center}
\end{figure}

As a further illustration, we present the complete evolution of the \ion{O}{i} line at 7773.0~\AA\ during
the whole light curve variation phase covering the observed spectra of the first 100 days after discovery.
Fig. \ref{fig:OI_2d} shows the variations in expansion velocities of the respective absorption features.
The variations in the light curve can be directly related to these changes in the expansion velocities.
Fig. \ref{fig:OI_2d} is also a great visualization of that the Nova V5668~Sgr shows all the classical systems
of spectra, which are principal, diffuse enhanced, and Orion \citep{mclaughlin43,mclaughlin44,kuiper1960stars}.

\begin{figure}
\begin{center}
 \resizebox{\hsize}{!}{\includegraphics{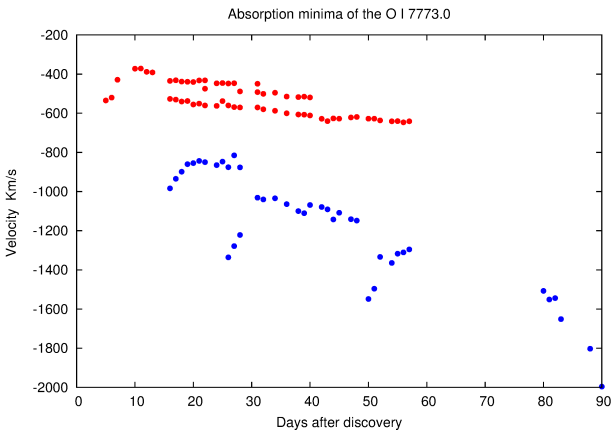}}
\caption{Evolution of the expansion velocities of the minima of the absorption features of the \ion{O}{i} line 
at 7773.0~\AA. There are two groups of minima one at around 500~km~s$^{-1}$ and the other between
1000 to 2000~km~s$^{-1}$.}
\label{fig:vel}
\end{center}
\end{figure}
To further illustrate and quantify the behaviour of the absorption features,
we measured the expansion velocity of the respective minima
for each day in the \ion{O}{i} line at 7773.0~\AA.
We used this line because it showed a simpler structure than the hydrogen lines,
which have various subfeatures and on some days many more minima.
In Fig.~\ref{fig:vel}, we show a graph of the position of all of the minima of
the absorption features of the \ion{O}{i} line.
There is one group of minima at around $-500$~km~s$^{-1}$ (red circles) which slowly
changes position to higher negative expansion velocity. This group of
absorption features disappears later, but because of an observation gap
we cannot determine on which day this happens.
The other group of absorption features can be found between 
expansion velocities of $-1000$ and $-2000$~km~s$^{-1}$ (blue circles).
While the slow expansion velocity group does not display big changes, the group
of higher velocity illustrates quite well the changes that occur
during a change in the visual light curve as can be seen in Fig. \ref{fig:lc}.
Comparing the graph of the absorption feature minima with the visual light curve,
one finds a good agreement. When the light curve is in a minimum,
the velocity of the absorption feature minimum also jumps to higher negative expansion
velocities.
During the following rise in the light curve this absorption feature moves back
to lower negative velocities.
The overall trend shows that while the visual
light curve slightly decreases during the variation phase of 90 days,
the absorption features also move to higher negative expansion velocities.
When the light curve eventually shows the steep decline towards
the deep minimum, the expansion velocities of the
absorption feature also moves to significantly higher negative
expansion velocities.

\subsubsection{Appearance of \ion{He}{i} features}

The list of lines that have features, which show the characteristic behaviour of Table \ref{tb:list},
contains two \ion{He}{i} lines at wavelengths of 6678.2 and 7065.2~\AA. These two
lines were not observed during the maximum before the second declining phase.
Instead, they appear in emission and show the high expansion velocity absorption
component only during the secondary minimum phase. This means that the
respective \ion{He}{i} lines get only excited and visible during this second minimum phase.

During the third minimum phase, the absorption features of these lines were already
observed during the maximum before and show the same shift in expansion velocity
as the features of the other lines. 
In the few spectra that we could obtain during the fourth declining phase,
we observed the same behaviour. During the fifth declining phase
the lines already show absorption features during the maximum, which then
shift during the declining phase to higher expansion velocities.

\subsubsection{Disappearance of \ion{Fe}{i} features}

\begin{figure}
\begin{center}
 \resizebox{\hsize}{!}{\includegraphics{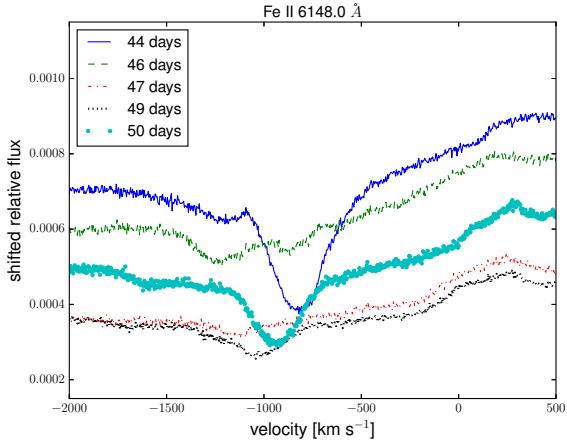}}
\caption{Evolution of the absorption features of the \ion{Fe}{ii} line at 6148.0~\AA\ during
the third minimum phase of the visual light curve of Nova V5668~Sgr. The features
disappear for a few days. The spectra are shown at different days after discovery.}
\label{fig:FeII}
\end{center}
\end{figure}

During the third declining phase of the visual light curve of 
Nova V5668~Sgr, the features of the \ion{Fe}{ii} line at a wavelength of 6148.0~\AA\ show
a very interesting behaviour in its spectral evolution, which is demonstrated in Fig. \ref{fig:FeII}.
During the third maximum 44 days after discovery, clear absorption features
at an expansion velocity of $\approx -800$~km~s$^{-1}$ are observed in the spectrum.
During the following declining phase (46 and 47 days after discovery) these features disappear.
However, 49 days after discovery they start to appear again, now at higher expansion velocities
of $\approx-1000$~km~s$^{-1}$. In the spectrum 50 days after discovery the absorption features become
more clearly visible. They also shift back to lower expansion
velocities, reaching almost their original position in the spectrum
51 days after discovery. Basically, the expansion velocity evolution observed in the features
of this \ion{Fe}{ii} line is very similar to those observed in other lines.
However, it is interesting that this \ion{Fe}{ii} line disappears for a few days,
while the light curve is in the declining phase.
Explanations may be that the \ion{Fe}{ii} might be
ionized to \ion{Fe}{iii} for a few days or that one just sees different layers with different
abundances of the expanding envelope.

Actually, absorption features of several \ion{Fe}{ii} lines disappear during the third
declining phase of Nova V5668~Sgr. Features of the \ion{Fe}{ii} line at a wavelength of 4232.0~\AA\ that
were observed during the third maximum disappear during the third declining phase.
In addition, there are several lines of \ion{Fe}{ii}
in the wavelength range between 5000 and 5300~\AA, 
i. e. 4232.5, 5018.4, 5169.0, 5234.6 and 5276.0~\AA, showing the same behaviour.
We also found a small emission feature at a wavelength of $\approx 6090$~\AA\ 
appearing during exactly this phase.

Although we only have one spectrum during the fourth declining phase,
we found that again some of the \ion{Fe}{ii} features observed during the maximum phase
disappear in the spectrum 61 days after discovery during the fourth declining phase.
Namely, these are the absorption features of the \ion{Fe}{ii} lines at wavelengths of 4491.4, 5316.6 and 6148.0~\AA.
Additionally, we observed that the absorption features of the \ion{Si}{ii} line at a wavelength of 6347.1~\AA\ disappear
during the fourth declining phase. During the fifth declining phase the absorption features
of the \ion{Fe}{ii} lines at 4491.4 and 6148.0~\AA\ do finally disappear.

\subsubsection{Disappearance of \ion{N}{i}, appearance of \ion{N}{ii}}

We found also an interesting behaviour of the absorption features
of the nitrogen lines in the spectra of Nova V5668 Sgr in the
light curve variation phase.
During the declining phase towards the third minimum absorption features of two \ion{N}{i} lines 
at wavelengths of 7442.3 and 7468.2~\AA\ disappear.
At the same time small emission features of \ion{N}{ii} lines appear at wavelengths of 3995.0,
5935.0 and 6483.8~\AA.
During the fourth declining phase the \ion{N}{i} lines at wavelengths of 7442.3 and 7468.2~\AA\ disappear
in the spectrum observed 61 days after discovery.
At the same time emission features of two
\ion{N}{ii} lines appear at wavelengths of 3995.0 and 6483.8~\AA.
During the fifth declining phase, emission features of the \ion{N}{ii} line at a wavelength of 3995.0~\AA\ appear
in the observed spectra of Nova V5668~Sgr.

\subsubsection{The forbidden [\ion{O}{i}] 5577.3~\AA\ line}\label{sec:forbid}

The only forbidden line that appears in the early observed spectra of Nova V5668~Sgr is
the one of [\ion{O}{i}] at a wavelength of 5577.3~\AA.
The first time it is visible is in the spectrum 9 days after discovery during the first declining phase.
The line was observed in emission, and the flux increased during the declining phases relatively to the 
continuum flux. However, this may be explained by a decreasing continuum flux as already noted above. 
During the following evolution of the visual light
curve during the variation phase, we see the same behaviour. During a minimum phase the emission feature of the [\ion{O}{i}]
line is clearly visible, while during a maximum phase it seems to disappear (in comparison to the continuum flux).
In the later spectra there appear features of other forbidden [\ion{O}{i}] lines at wavelengths of
6300.3 and 6363.7~\AA\ as can be seen in Fig.\ref{fig:all_specs_R}.

\subsubsection{Daily variations in the \ion{Na}{i} D doublet line}

\begin{figure}
\begin{center}
 \resizebox{\hsize}{!}{\includegraphics{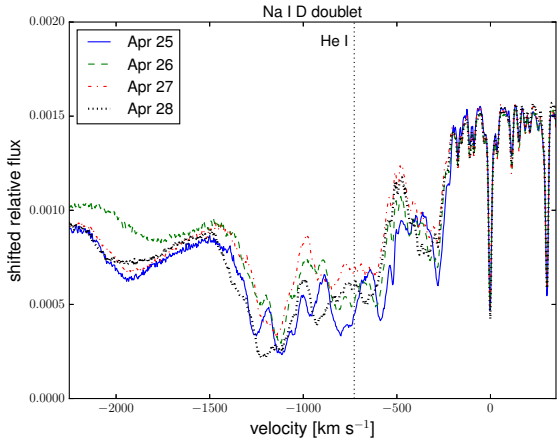}}
\caption{Daily variations observed in the absorption features
of the \ion{Na}{i} doublet line in the spectra of Nova V5668~Sgr.}
\label{fig:Na}
\end{center}
\end{figure}

From April 25 to April 28 in 2015, which corresponds to 41 to 44 days after discovery,
we were able to obtain a spectrum of Nova V5668~Sgr each
night. Comparing these spectra we found that the features of the \ion{Na}{i} D doublet lines
at wavelengths of 5895.9 and 5889.9~\AA\ show daily variations.
In Fig. \ref{fig:Na}, we present the corresponding spectra in the wavelength range of
the \ion{Na}{i} doublet, where we took the 5889.9~\AA\ line as the reference for the velocities.
The two strong narrow lines at the right hand part of the graph are features from interstellar absorption
of both \ion{Na}{i} lines.
Several absorption features of the expanding nova envelope can be seen in the
blue part of the \ion{Na}{i} doublet lines.
The structure of the features are changing every day.
It is difficult to quantify this evolution since the features of both lines
overlap and therefore can cause the daily changes. There is also the above mentioned
\ion{He}{i} line at 5875.6~\AA, which might have absorption and/or emission features that could
contribute to the flux in that wavelength region and, therefore, affect the daily changes.
The position of this \ion{He}{i} line has been marked in the spectra shown in Fig. \ref{fig:Na}.

\subsection{Emission line phase}
\begin{figure}
\begin{center}
 \resizebox{\hsize}{!}{\includegraphics{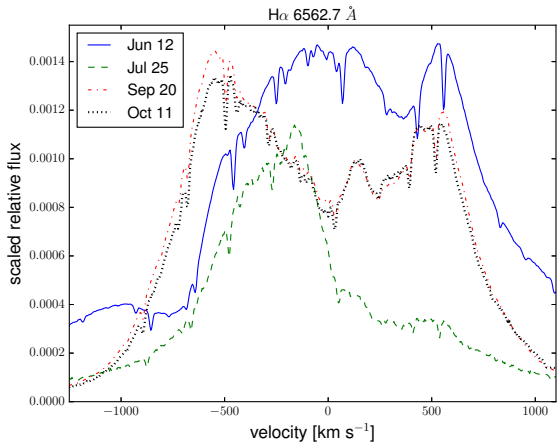}}
\caption{Evolution of the emission features of the H~$\alpha$ line at 6562.7~\AA\ in
the spectra of Nova V5668~Sgr during the later phases.}
\label{fig:Halpha}
\end{center}
\end{figure}

After recovering from the deep minimum in the light curve
all lines show only emission features in our spectra of Nova V5668 Sgr, meaning that
the nova is now in its nebular phase.
During the rising phase after the minimum the visual light curve of the nova (Fig. \ref{fig:lc_tot}) shows
some fluctuations, e.g. at around 150 days after discovery.
However, our observed spectra show for all lines a simple one peak emission line feature,
but that is probably because of the low signal-to-noise ratio of the
spectra during that phase, since the nova was still too dim for good observations with the TIGRE telescope.
The only line showing more complex features is H~$\alpha$ due to being the by far strongest line.
In Fig. \ref{fig:Halpha}, we show the evolution of the shape of the emission features of the H~$\alpha$ line
at 6562.7~\AA.
The spectrum from June 12, 89 days after discovery, observed during the steep decline
towards the deep minimum, shows an emission feature which has a maximum close to 0~km~s$^{-1}$. The feature is almost symmetrical, but
it shows a small peak around $+600$~km~s$^{-1}$ and a dip around $-750$~km~s$^{-1}$. 

The narrow absorption features that can be seen in all of the spectra are due to telluric lines.
The emission feature of the H~$\alpha$ line in the spectrum from July 25, 132 days after discovery,
when the light curve is rising again,
is strong on the blue side with negative expansion velocity,
while on the positive expansion velocity side only a small emission is observed. The maximum
of the emission feature can be found around $-200$~km~s$^{-1}$.
When the light curve is in the almost constant phase after the deep minimum,
the H~$\alpha$ emission feature of Nova V5668 Sgr has a clear double peak structure.
This can be seen in the observed spectrum from September 20 (189 days after discover) as well
as in the spectrum from October 11 (210 days after discovery). The peak on the negative expansion velocity side
is a bit higher than the second peak on the other side. Comparing the spectra from September 20 with
the one from October 11, one can see that there are hardly any changes in the shape of the emission feature
of the H~$\alpha$ line.

\section{Conclusions and discussion}\label{sec:conclusion}

With the TIGRE robotic telescope,
we obtained a dense time series of high resolution spectra ($R\approx 20,000$)
of Nova V5668~Sgr, which was discovered on March 15 in 2015,
in the optical wavelength range between 3800 and 8800~\AA.
The nova is a typical DQ Her type nova,
and it shows all the known classical systems of spectra:
principal, diffuse enhanced, and Orion spectra.
We used the AAVSO visual light curve of Nova V5668~Sgr to spectroscopically characterise
different phases, beginning with a phase of strong variations followed by a deep,
prolonged minimum, leading then to a recovery and a phase at an almost constant brightness 
level. We could place 
Nova V5668~Sgr at an estimated distance of 1.6~kpc assuming a moderate extinction
and a characteristic absolute magnitude of a classical nova type of DQ Her. 

We presented four characteristic spectra of Nova V5668~Sgr and
performed a detailed line identification on these
respective spectra. We find that all the Hydrogen Balmer lines are
present in all of the observed spectra, with H~$\alpha$ the strongest line.
The nova also shows many lines of iron in form of \ion{Fe}{ii} and oxygen
mainly in form of \ion{O}{i}. In addition, we found some lines of
\ion{Si}{ii}, \ion{Ti}{ii}, \ion{N}{ii}, \ion{Mg}{ii}, and \ion{Na}{i} among others. In the
later spectra we find some forbidden lines of oxygen ([\ion{O}{i}] and [\ion{O}{ii}]).
During the light curve evolution, the profiles of the features
change from being P-Cygni profiles to later having only emission features.

We studied the observed spectra in order to find a correlation with
the variations observed during the first 90 days in the visual light curve of Nova V5668~Sgr.
We find that it is mainly the continuum flux which varies during the changes in the visual 
light curve. We can distinguish five relatively steep declines in the visual light curve of the nova.
Our dense series of spectra reveals that during the fast declining phases the absorption features
of many lines, especially the ones of hydrogen, shift to higher expansion velocities
of about $-2000$ to $-1500$~km~s$^{-1}$. During the following, generally slower rises of the light curve,
these absorption features shift back to lower expansion velocities. 

We also find that some absorption features, such as those of 
\ion{Fe}{ii}, disappear for a few days during the third, fourth and 
fifth declining phases of the visual light curve, as well as 
one \ion{Si}{ii} line during the fourth declining phase. 
During the third declining phase, new \ion{He}{i} lines appear in 
the spectra, which had not been recorded before.
In addition, we find that absorption features of some \ion{N}{i} lines disappear while new
emission features of \ion{N}{ii} lines appear in the spectra during the third, fourth, and
fifth declining phases in the light curve.
The \ion{Na}{i} D doublet line at around 5900~\AA\ has many features
that show daily variations during the first 90 days after discovery, which are probably
due to a blend of the absorption features of the two \ion{Na}{i} lines with the \ion{He}{i} line at 5875.6~\AA.

Since a nova has a typical radial wind profile of increasing velocity with increasing 
radius, one might suggest that what is seen during the minimum phases are layers of
the envelope that lie farther outside and, therefore, show higher expansion velocities,
occulting the layers and their less shifted line contributions from below.
Such an optical depth effect would also explain the synchronisation of these changes in 
several, different lines:
Comparing a graph of the positions of the minima of the absorption features
of the \ion{O}{i} line at 7773.0~\AA, we find a good agreement with
the visual light curve. Whenever the light curve decreases the absorption
features move to higher expansion velocities.
The second group of absorption features around $-500$~km~s$^{-1}$, which are
also visible in many other lines, only shows
the general trend of moving towards higher negative expansion velocities
while the light curve is decreasing. However, the strong appearance of the
highly blue shifted absorption just after any decline may also be explained by 
real, temporary changes in the expansion velocity, synchronized by some physical process
with the steep brightness decreases of the nova. The same process may then also
explain the synchronized changes in ionisation reported above. In any case, a lot of 
further analysis will be required to investigate these or other possible interpretations 
of the rich spectroscopic evidence presented here.

After having been enshrouded by dust \citep{ATel7643, ATel7748} since around 90 days
after discovery, the light curve of Nova V5668~Sgr has started to rise again around 130 days
after discovery. The nova brightness having then returned to about 9 mag,
we were able to resume our spectroscopic monitoring. After the deep minimum the lines in
the spectra show only emission features and no absorption features.
The emission features have a clear double peak structure
as can be clearly seen in the strongest line of H~$\alpha$ at 6562.7~\AA. The peak
on the blue side is a bit higher than the red one.

The unique data set obtained by our ongoing dense spectroscopic monitoring of
Nova V5668~Sgr will be further analysed in a future publication, which will include 
detailed modelling of some spectra and will give us the opportunity to
study the spectral evolution of this DQ Her type nova in much more detail.

\acknowledgements
We acknowledge with thanks the variable star observations
from the AAVSO International Database contributed by
observers worldwide and used in this research.
Our collaboration and work was much helped by travel money from 
bilateral (Conacyt-DAAD) project grant No. 207772 as well as 
Conacyt mobility grant No. 207662.
We kindly acknowledge the DFG Graduiertenkolleg GrK 1351 for funding support
as well as funding of the DAIP-UG (534/2015) project grant of the University of Guanajuato.

\bibliographystyle{an}
\bibliography{all}

\label{lastpage}
\end{document}